\documentclass[%
singlecolumn,
superscriptaddress,
amsmath,amssymb,
aps,
showpacs,
]{revtex4-1}
\usepackage{boites}
\usepackage{amsmath}
\usepackage{makeidx}
\usepackage{float}
\usepackage{graphicx,color}
\usepackage{amsfonts}
\usepackage[usenames,dvipsnames]{pstricks}
\usepackage{subfigure}
\usepackage{epsfig}
\usepackage{pst-grad}
\usepackage{pst-plot}
\usepackage{mathrsfs}

\makeindex            
 
\def\vect#1{\mbox{\boldmath $#1$}}
\newcommand{\oog}{${{^{16}}{\rm O}}$}
\newcommand{\neneg}{${^{20}{\rm Ne}}$}
\newcommand{\ccg}{${^{12}{\rm C}}$}

\def\13C{{^{13}{\rm C}}}
\begin{document}
	\title{Nonlocalized motion in two-dimensional container of $\alpha$ particles in $3^-$ and $4^-$ states of $^{12}$C}
	\author{Bo Zhou}
	\affiliation{Institute for the Advancement of Higher Education, Hokkaido University, Sapporo 060-0817, Japan}
	\affiliation{Department of Physics, Hokkaido University, 060-0810 Sapporo, Japan}
\author{Yasuro Funaki}
\affiliation{College of Science and Engineering, Kanto Gakuin University, Yokohama 236-8501, Japan}
\author{Hisashi Horiuchi}
\affiliation {Research Center for Nuclear Physics (RCNP), Osaka University, Osaka  567-0047, Japan}
\author{Masaaki Kimura}
\affiliation{Department of Physics, Hokkaido University, 060-0810 Sapporo, Japan}
\affiliation{Reaction Nuclear Data Centre, Faculty of Science, Hokkaido University, 060-0810 Sapporo, Japan}
\author{Zhongzhou Ren}
\affiliation{School of Physics Science and Engineering, Tongji University, Shanghai 200092, China}
\author{Gerd R\"{o}pke}
\affiliation {Institut f\"{u}r Physik, Universit\"{a}t Rostock, D-18051 Rostock, Germany}
\author{Peter Schuck}
\affiliation{Institut de Physique Nucl\'{e}aire, Universit\'e Paris-Sud, IN2P3-CNRS, UMR 8608, F-91406, Orsay, France}
\author{Akihiro Tohsaki}
\affiliation{Research Center for Nuclear Physics (RCNP), Osaka University, Osaka 567-0047, Japan}
\author{Chang Xu}
\affiliation{Department of Physics, Nanjing University, Nanjing 210093, China}
\author{Taiichi Yamada}
\affiliation{Laboratory of Physics, Kanto Gakuin University, Yokohama 236-8501, Japan}
	\date{\today}
\begin{abstract}
The first $3^-$ and $4^-$ states of $^{12}$C are studied in the present container model, in which the shift parameter is introduced to break the parity symmetry for projecting out the negative-parity states. Taking the limit as the shift parameter approaches zero and by variational calculations for one-deformed size parameter, the local energy minima are obtained for the $3^-$ and $4^-$ states. 
It is found that the obtained {\it single} THSR (Tohsaki-Horiuchi-Schuck-R\"{o}pke) wave functions for $3^-$ and $4^-$ states are 96\% and 92\% equivalent to the corresponding GCM wave functions, respectively. 
The calculated intrinsic densities further show that these negative-parity states of three clusters, different with the traditional understanding of rigid triangle structure, are found to have nonlocalized clustering structure in the two-dimensional container picture.
\end{abstract}
	\maketitle
The \ccg\ is a specially interesting atomic nucleus due to its rich cluster structure which has been investigated for half a century by various models~\cite{horiuchi_manycluster_1975,uegaki_structure_1977,kamimura_transition_1981,navratil_properties_2000,epelbaum_structure_2012,freer_hoyle_2014,funaki_hoyle_2015,kanada-enyo_isoscalar_2016,freer_microscopic_2018}. In particular, describing and understanding the essential correlations of 3$\alpha$-cluster states is a fundamental problem in this nucleus. Recently, using the THSR wave function~\cite{tohsaki_alpha_2001,tohsaki_colloquium_2017}, the Hoyle state was proposed to have a three $\alpha$ condensate characters where the 3$\alpha$ clusters have 70\%$-$80\% possibility of making relative $(0S)$ motion~\cite{yamada_single_2005} in \ccg.

Quite recently, inspired by the original THSR wave function, we proposed a container model~\cite{zhou_nonlocalized_2013,zhou_nonlocalized_2014,zhou_container_2014,funaki_container_2018} for the description of general cluster states in nuclear systems. Different from the traditional understanding, the clusters are making a nonlocalized motion in the container which is only confined by the size parameter instead of the inter-cluster separation-coordinates. One unique feature of this new model is that we  can obtain with high-precision a {\it single} optimal wave functions for various cluster states. It enables us to discuss the correlations of clusters in a straightforward way based on the explicit relative wave functions. This was highlighted by the successful descriptions of many nuclear systems~\cite{lyu_investigation_2016,zhou_breathinglike_2016,zhao_investigation_2018,funaki_container_2018}.

However, up to now, the important negative-parity cluster states in $^{12}$C still have not been studied in the container model due to the fact that the THSR wave function in its original form described only positive-parity states. Traditionally, in analogy with the usual belief of the typical rigid $\alpha$+\oog\ structure in \neneg\ system~\cite{horiuchi_moleculelike_1968}, the structure of the $^{12}$C has also long thought to be associated with a definite triangular shape. One typical state is the $3^-$ state at 9.64 MeV, which could have threefold symmetry and three alpha particles are in an equilateral triangular arrangement. Recently, Bijker et al.~\cite{bijker_algebraic_2002,bijker_algebraic_2017} proposed a description of cluster states in nuclei in terms of representations of unitary algebras. The $^{12}$C was assumed to have $D_{3h}$ 3$\alpha$ cluster structure. Based on the  algebraic cluster model, the rotational band of an oblate equilateral triangular with the cluster states $0^+, 2^+, 3^-, 4^{\pm}, 5^-$ was predicted. The negative-parity states was considered as the important support for their rigid 3$\alpha$ cluster picture in $^{12}$C. However, this model became also criticized of not respecting the Pauli principle~\cite{hess_12c_2018} or the positive and negative states can be interpreted as two separate bands~\cite{cseh_symmetries_2018}. Quite recently, a nuclear fluorescence experiment through patterns of polarized $\gamma$ rays was proposed~\cite{fortunato_establishing_2019} to discriminate the geometric configurations for $\alpha$-cluster nuclei like \ccg\ in a model-independent way. On the other hand, the THSR wave function has taught us that the positive-parity states, especially the Hoyle state, should not be seen as being in rigid geometrical 3$\alpha$ arrangements but rather the $\alpha$'s are in a low-density gas state akin to a Bose condensate. Thus, it is highly desirable to study the negative-parity states in the container model and clarify the corresponding structure of 3$\alpha$ clustering in a microscopic way using a generalized THSR wave function.

\begin{figure}[!h]
	\begin{center}
		\includegraphics[scale=0.63]{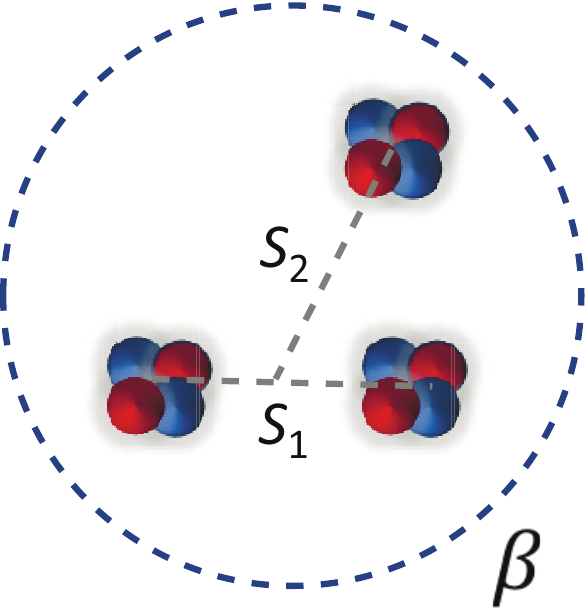}
		\caption{Schematic diagram for the container picture in the description of negative-parity cluster structure of \ccg.}
		\label{fig:container-neg}
	\end{center}
\end{figure}

To describe the negative-parity states, we here introduce  shift parameters to generate negative-parity states in the THSR wave function. This  has successfully been used already for the negative-parity states of the two-cluster structure in \neneg~\cite{zhou_nonlocalized_2013}. The simple THSR wave function for three $\alpha$'s with one size parameter $\vect{\beta}$ can be constructed as,  
\begin{align}
\Phi(\vect{\beta},\vect{S}_1,\vect{S}_2) &= \!\int\!d^3 R_1 d^3 R_2 \exp[-\frac{(\vect{R}_1-\vect{S}_1)^2 }{2\vect{\beta}^2}-\frac{2(\vect{R}_2-\vect{S}_2)^2 }{3\vect{\beta}^2} ] \Phi^B(\vect{R}_1,\vect{R}_2)  \nonumber  \\ 
 &\propto \phi_G  {\cal A} \{ \exp[-\frac{(\vect{\xi}_1-\vect{S}_1)^2}{B^2} -\frac{(\vect{\xi}_2-\vect{S}_2)^2}{3/4~B^2} \phi(\alpha_1)\phi(\alpha_2)\phi(\alpha_3)  ]  \},\label{hyb} \\
\Phi^B(\vect{R}_1,\vect{R}_2 ) 
&\propto \phi_G {\cal A} \{ \exp{(-\frac{(\vect{\xi}_1-\vect{R}_1)^2}{b^2}-\frac{(\vect{\xi}_2-\vect{R}_2)^2}{3/4~b^2})} \phi(\alpha_1)\phi(\alpha_2)\phi(\alpha_3) \}, \label{brink}
\end{align}
where  $B^2=b^2+2\vect{\beta}^2$.
The $^{12}$C is assumed to have a $z-$axial symmetry shape, i.e., $\vect{\beta}\equiv(\beta_{x}=\beta_{y},~\beta_{z})$. $\vect{\xi}_1=\vect{X}_2-\vect{X}_1$ and $\vect{\xi}_2=\vect{X}_3-(\vect{X}_1+\vect{X}_2)/2$. The $\vect{X}_1$, $\vect{X}_2$, and $\vect{X}_3$ are the center-of-mass coordinates of the three clusters of \ccg. In the above equation, $\Phi^B(\vect{R}_1,\vect{R}_2 )$ is the 3$\alpha$  Brink wave function~\cite{brink_proceedings_1966} with the corresponding generator coordinates $ \vect{R}_1$ and $\vect{R}_2$.
The $ \vect{R}_1$ represents the specified distance parameter of two clusters and $\vect{R}_2$ the distance between the center-of-mass of these two clusters and the third $\alpha$ cluster. The shift parameters $ \vect{S}_1$ and $\vect{S}_2$ are introduced  to deal with the negative-parity states. Their coordinates,  $\vect{S}_1\equiv(S_{1x}, 0,0)$ and  $\vect{S}_2\equiv(S_{2x},S_{2y},0)$,  can arrange arbitrary triangular shapes in the $xy-$plane as shown in Fig.\ref{fig:container-neg}. Thus, applying angular momentum and parity projection techniques, we can obtain negative-parity components from the extended THSR wave function. It should be noted that the above extended THSR wave function can also be considered as a Hybrid-THSR-Brink wave function because there are shift {\it and} width parameters.  As for the Hamiltonian,  the Volkov No.2~\cite{volkov_equilibrium_1965} (modified version) with Majorana parameter $M$ = 0.59 and harmonic-oscillator size parameter $\nu$ =1/(2$b^2$)= 0.275 fm$^{-2}$ is used, which is the same as employed by Kamimura et al. for their 3$\alpha$ RGM calculations~\cite{kamimura_transition_1981}.

In practical calculations for the negative-parity states of \ccg, one important problem is how to set the proper shift parameters $\vect{S}_1$ and $\vect{S}_2$ in the three-cluster system as shown in Fig.~\ref{fig:container-neg}. The treatment of negative-parity states in \neneg\ told us that~\cite{zhou_nonlocalized_2013} the introduced shift parameter only plays the role of breaking the parity symmetry of the THSR wave function in an  intermediate step while the size parameter is the essential dynamical variable for the description of clustering motion.
In analogy with the \neneg\ case~\cite{zhou_nonlocalized_2013}, we slightly separate the 3$\alpha$ clusters to form an equilateral triangle shape on $xy-$plane, namely $\vect{S}_1=(S,0,0)$, $\vect{S}_2=(0,\sqrt{3}/2S,0)$, and take the side-length very small, i.e., $S$=0.5 fm. Note that the shift parameter $\vect{S}_1$ actually can be zero due to the 2$\alpha$ positive-parity character and the shift orientation can also be taken three dimensional. Here this kind of numerical  "infinitesimal shifted triangle" is our first simple try and later we will discuss our wave function with respect to  the dependence on shift parameters.

Based on this simple THSR wave function, we performed variational calculations after  angular-momentum and parity projections for the $3^-$ and $4^-$ states in the two parameter $\beta_{x}=\beta_{y}$ and $\beta_{z}$ space. Here we only consider the dominant $K$=3 component in the single projected THSR wave functions.
The reason why we only consider the $3^-$ and $4^-$ states and not also the $1^-$ and $5^-$ states has to do with the fact that the latter states apparently have a more complicated structure or resonance character than describable with the present ansatz of our wave function. It is planned to devote an extra study for those states by superposing THSR wave functions and also applying the resonance method in the future. Here we concentrate on the first two states which are well describable with our technique as it will be demonstrated below.

\begin{figure}[!htb]
	\begin{center}
		\includegraphics[scale=0.46]{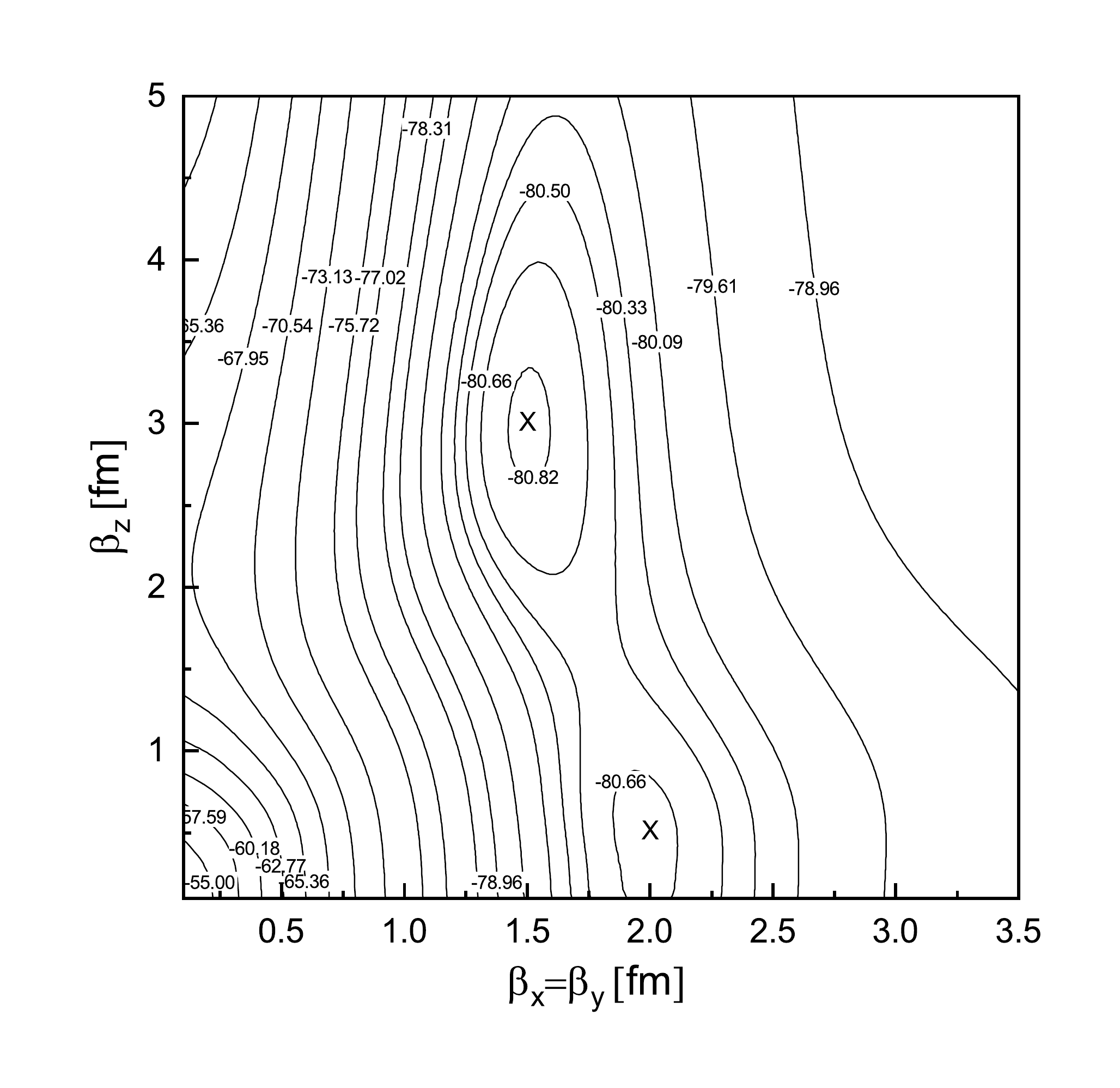}
		\caption{Energy contour plot for the $3^{-}$ state in the two-parameter space $\beta_x=\beta_y$ and  $\beta_z$ in the container model.}
		\label{fig:neg3k3sxy05pic}
	\end{center}
\end{figure}

\begin{figure}[!htb]
	\begin{center}
		\includegraphics[scale=0.46]{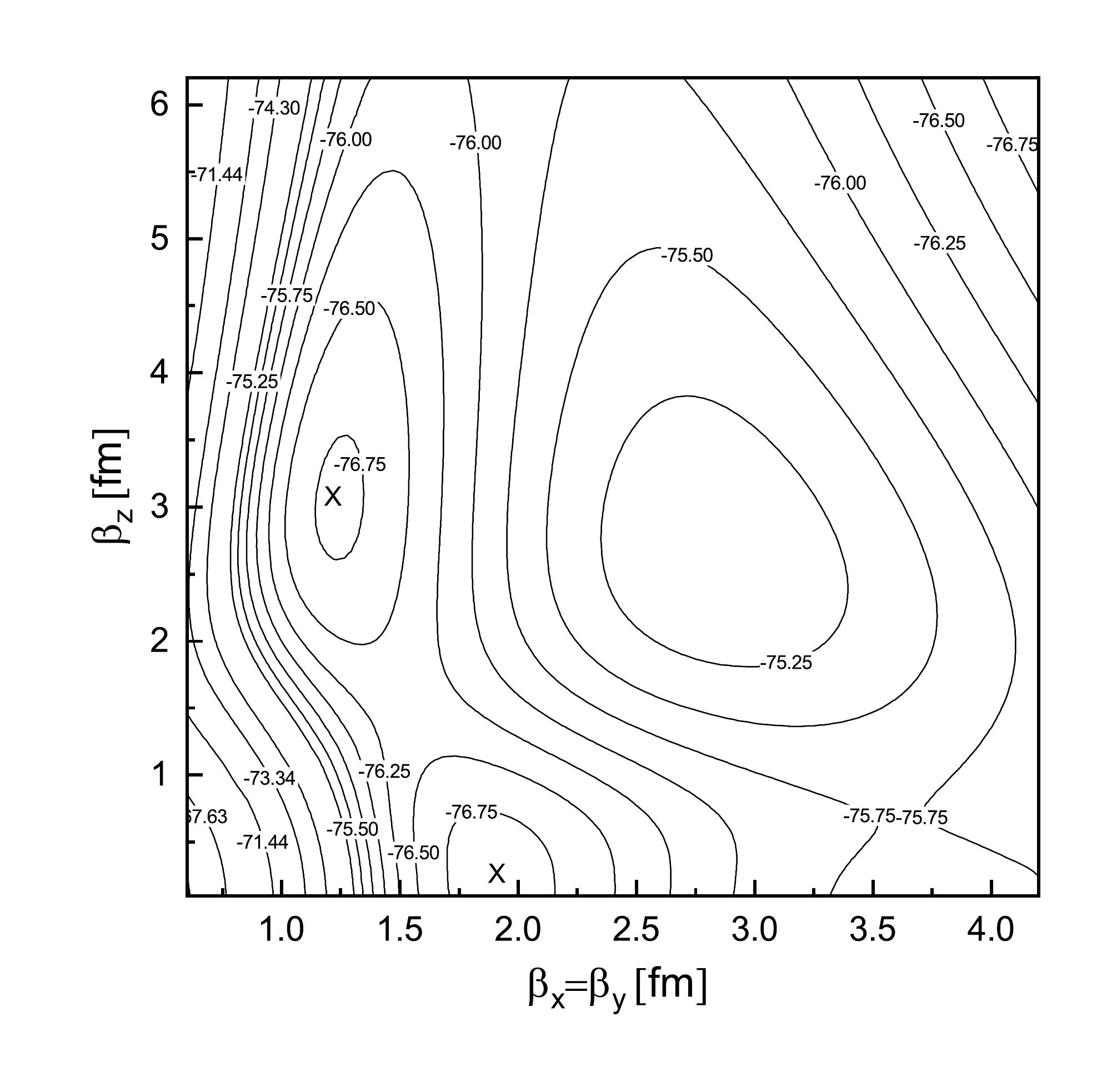}
		\caption{Energy contour plot for the $4^{-}$ state in the two-parameter space $\beta_x=\beta_y$ and  $\beta_z$ in the container model.}
		\label{fig:neg4k3sxy05pic}
	\end{center}
\end{figure}

Figure~\ref{fig:neg3k3sxy05pic} shows the energy contour plot for the $3^{-}$ state after variational calculations with respect to the $\vect{\beta}$ parameters. The value for the side-length of the shift equilateral-triangle is taken 0.5 fm as mentioned. It can be seen that  two local minimum points have appeared, which are connected by a long and narrow valley. The deeper local minimum point appears at $\beta_x=\beta_y$=1.5 fm and $\beta_z$=3.0 fm. The obtained corresponding energy for the  $3^{-}$ state is about $-$80.9 MeV, which is very close to the result of the previous RGM calculation $-$81.2 MeV by Kamimura {\it et al.}~\cite{kamimura_transition_1981}. The second local minimum point appears at $\beta_x=\beta_y$=2.0 fm and $\beta_z$=0.5 fm, whose energy is about $-$80.7 MeV. The wave functions of these two local minimum points are quite similar and their squared overlap is about 98\%. Figure~\ref{fig:neg4k3sxy05pic} shows the contour plot for the  $4^{-}$ state. We also can find two local energy minima and the deeper one appears at $\beta_x=\beta_y$=1.9 fm and $\beta_z$=0.2 fm, whose energy is $-$76.9 MeV. Another local minimum point appears at $\beta_x=\beta_y$=1.2 fm and $\beta_z$=3.0 fm. The value of the squared overlap between these two obtained optimal wave functions with prolate and oblate shapes is also as high as 98\%. It can be seen that, this kind of equivalence of projected prolate and oblate THSR wave functions is very similar to the \neneg\ case. This actually has its reason in the nonlocalized character of the THSR wave function which has been discussed in the Ref.~\cite{zhou_nonlocalized_2014}. It also should be noted that the $3^{-}$ and $4^{-}$ states are shown to have quite similar oblate intrinsic shapes ($\beta_x=\beta_y$$\approx$2.0 fm and $\beta_z$$\approx$0.5 fm) and we will see later that this character is very useful for our analysis of the intrinsic shapes.

\begin{figure}[!htb]
	\begin{center}
		\includegraphics[scale=0.44]{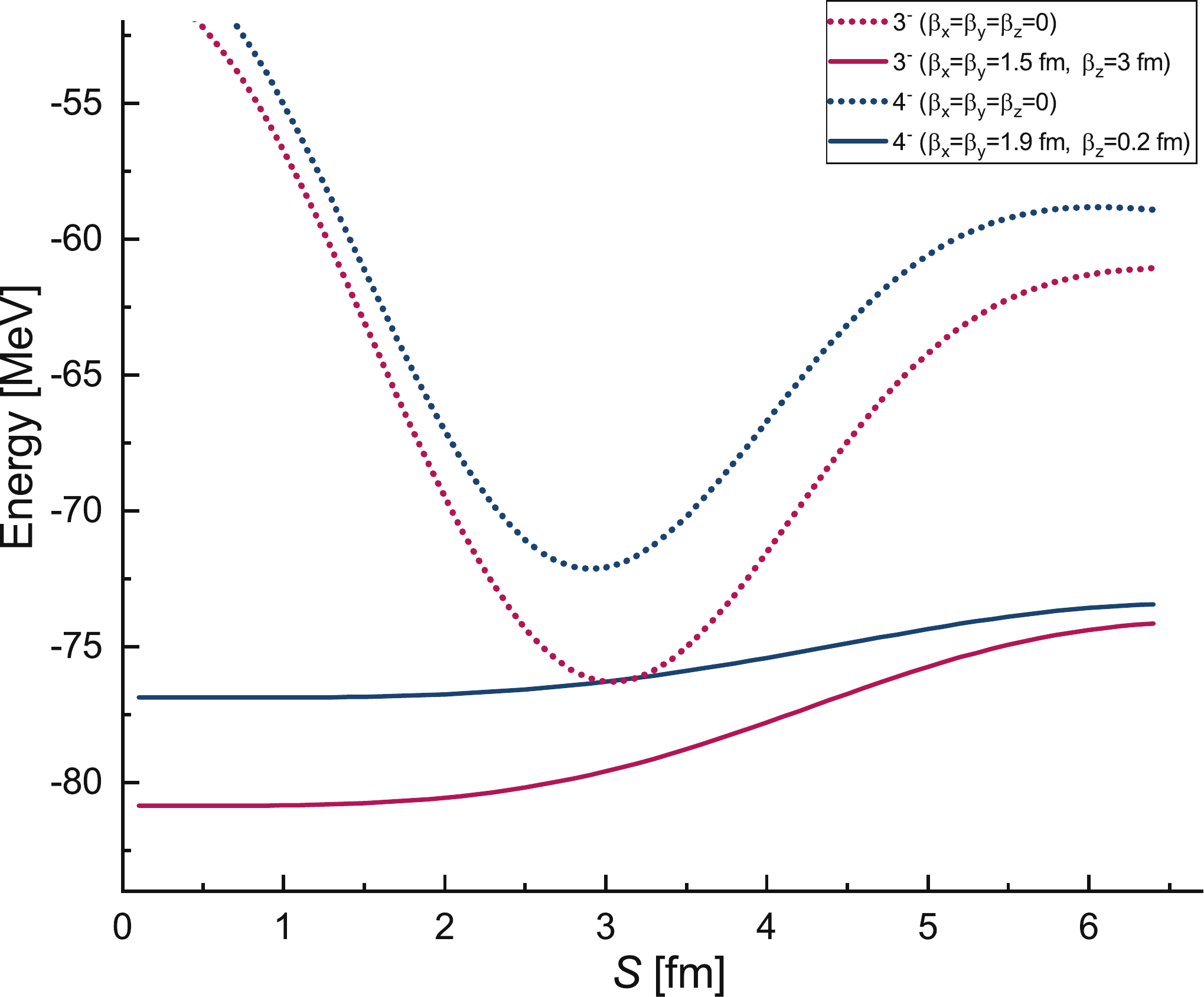}
		\caption{Energy curves (calculated from the Eq.~(\ref{hyb}) and see the text.) of the $3^-$ and $4^-$ states of \ccg\ within the THSR wave functions compared with those from the Brink wave functions.}
		\label{fig:negnonloc}
	\end{center}
\end{figure}

\begin{table}[!htb]
	\centering
	\caption{The calculated energies from the single optimal THSR wave functions in Eq.~(\ref{hyb}), the single optimal
	Brink wave functions in Eq.~(\ref{brink}), and the Brink-GCM  wave functions for the $3^-$ and $4^-$ states. The values of the squared overlap between the single optimal THSR/Brink wave functions and the Brink-GCM wave functions are also shown.}
	\begin{tabular}{ c c c c c c c}
	\hline
	\hline
	$J^\pi$&$E^{\text{\tiny{Brink}}}_{\text{ min}}(\vect{R}_1,\vect{R}_2)$ & $E^{\text{\tiny{THSR}}}_{ \text{min}} (\vect{\beta})$& $ E_{\text{\tiny GCM}}^{\text{\tiny{Brink}}}$ &
	$ |\langle \Phi_{\text{\tiny{GCM}}}^{\text{\tiny{Brink}}}| \Phi^{\text{\tiny{Brink}}}_{\text{ min}}(\vect{R}_1,\vect{R}_2)\rangle|^2 $  & $|\langle \Phi_{\text{\tiny{GCM}}}^{\text{\tiny{Brink}}}|\Phi^{\text{\tiny{THSR}}}_{\text{ min}}(\vect{\beta})\rangle|^2$  \\ \hline
	$3^-$ & $-78.4$&$-80.9$&$-81.6$&$0.78$&$0.96$    \\
	$4^-$ & $-74.4$&$-76.9$&$-77.8$&$0.72$&$0.92$  \\  
	\hline
	\hline
	    \label{tab}
\end{tabular}
\end{table}

In Fig.~\ref{fig:negnonloc}, we compare the energy curves obtained from the THSR wave functions with those from the Brink cluster model. At first, we assume the equilateral triangle shape with the length of side $S$ for 3$\alpha$ clusters in the Brink model, which is just equivalent to $\vect{\beta} \to$ 0 for the extended THSR wave function in Eq.~(\ref{hyb}). After variational calculations, it can be seen that there are two distinct pockets around 3 fm for the $3^-$ and $4^-$ states. It seems that, from this microscopic cluster model, there is a support for the rigid geometrical structure due to the non-zero inter-cluster distance parameter $S$. However, if we introduce the width variable of the relative wave function, i.e., the non-zero size parameter $\vect{\beta}$, the minimum energy points appear around $S \approx 0$.
This means the introduced $S$ parameter, like in the \neneg($\alpha$+\oog) case~\cite{zhou_nonlocalized_2013}, only plays an important role for treating the negative-parity states as a shift parameter in this three-cluster system. In particular, the energy curves from the THSR wave function are very flat within the range $S<2$ fm. The obtained energies are almost degenerate without depending on the inter-cluster distance parameter $S$. 
Furthermore, it can be seen that the obtained energies in the container model for  $3^-$ and $4^-$ states are 2.5 MeV deeper than those obtained from the Brink model. Moreover, we performed GCM calculations for  $3^-$ and $4^-$ states by superposing hundreds of projected Brink wave functions with different inter-cluster generator coordinates and $K$ numbers. The obtained GCM results are consistent with recent calculations from the real-time evolution method~\cite{imai_realtime_2018}. In Table~\ref{tab}, we list the squared overlaps between the single THSR wave functions and the GCM wave functions, which are as high as 96\% and 92\%  for the $3^-$ and $4^-$ states, respectively. On the other hand, the squared overlaps are only 78\% and 72\%  in the case of the single optimal Brink wave functions, which were obtained by variational calculations from arbitrarily shaped triangle configurations. This indicates that the single specific triangle configuration cannot reproduce well these negative-parity states while these cluster states can be described very well in the container picture. 

\begin{figure}[!htbp]
	\centering
	\subfigure[~~~~~~~~
	$|\langle\Phi^{3^-}\!(1/2,0,\sqrt{3}/4)|\Phi_{\text{\tiny{GCM}}}^{3^-} \rangle|^2$=0.95
	$|\langle\Phi^{4^-}\!(1/2,0,\sqrt{3}/4)|\Phi_{\text{\tiny{GCM}}}^{4^-} \rangle|^2$=0.93
	]{
		\begin{minipage}[t]{0.251\linewidth}
			\centering
			\includegraphics[width=1.3in,height=1.3in]{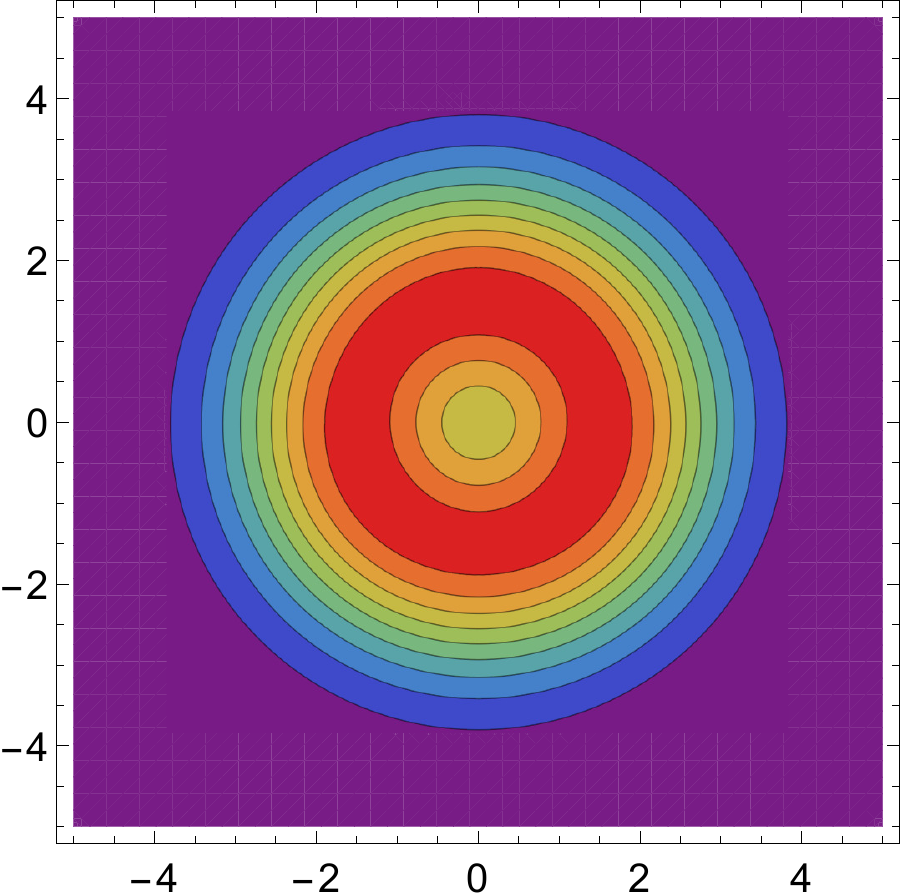}
			\label{figc1}
		\end{minipage}%
	}%
	\subfigure[~~~~~~~~
	$|\langle\Phi^{3^-}\!(3/2,3/2,1/2)|\Phi_{\text{\tiny{GCM}}}^{3^-} \rangle|^2$=0.94
	$|\langle\Phi^{4^-}\!(3/2,3/2,1/2)|\Phi_{\text{\tiny{GCM}}}^{4^-} \rangle|^2$=0.92
	]{
		\begin{minipage}[t]{0.251\linewidth}
			\centering
			\includegraphics[width=1.3in,height=1.3in]{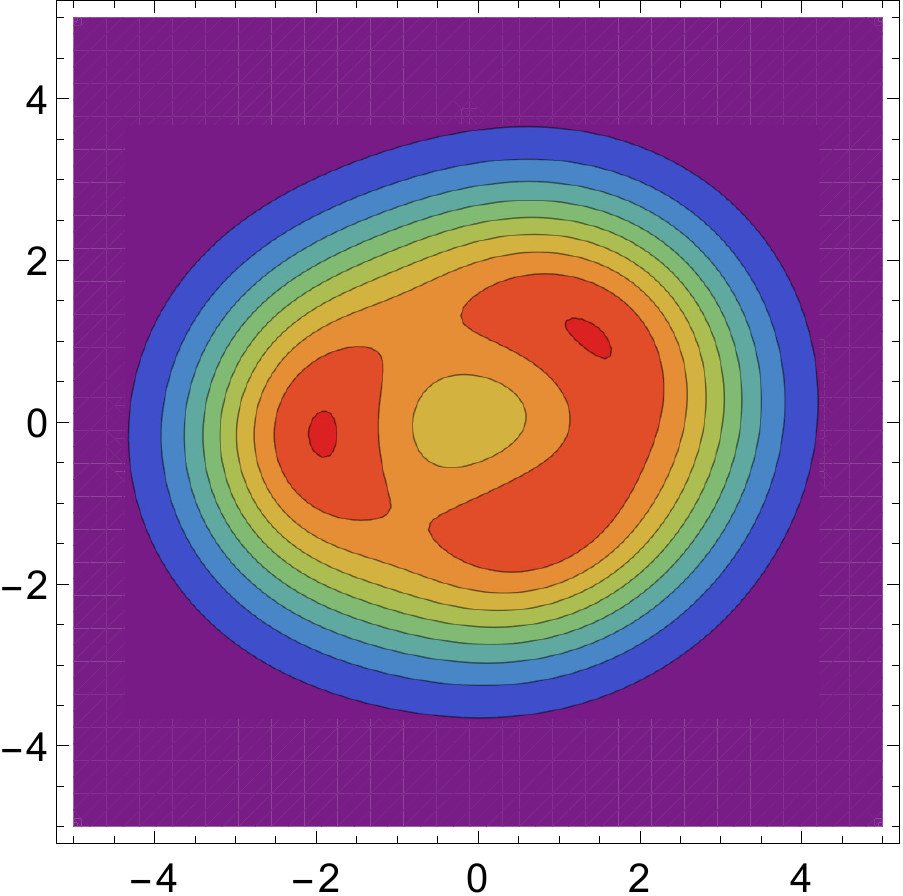}
		\end{minipage}%
	}%
	\subfigure[~~~~~~~~
	$|\langle\Phi^{3^-}\!(1,3/2,3/2)|\Phi_{\text{\tiny{GCM}}}^{3^-}\rangle|^2$=0.93
	$|\langle\Phi^{4^-}\!(1,3/2,3/2)|\Phi_{\text{\tiny{GCM}}}^{4^-}\rangle|^2$=0.92
	]{
		\begin{minipage}[t]{0.245\linewidth}
			\centering
			\includegraphics[width=1.3in,height=1.3in]{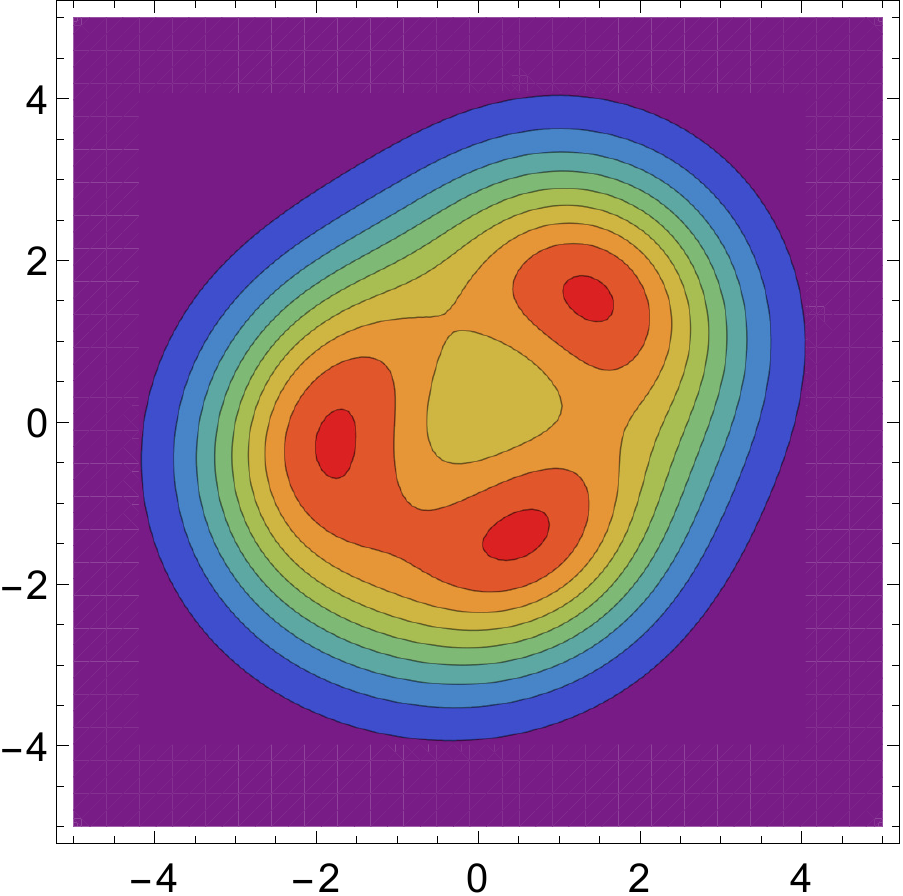}
		\end{minipage}
	}%
	\subfigure[~~~~~~~~~~~~~~~~~~~
	$|\langle\Phi^{3^-}\!(3/2,0,3/2)|\Phi_{\text{\tiny{GCM}}}^{3^-} \rangle|^2$=0.94
	$|\langle\Phi^{4^-}\!(3/2,0,3/2)|\Phi_{\text{\tiny{GCM}}}^{4^-} \rangle|^2$=0.92
	]{
		\begin{minipage}[t]{0.251\linewidth}
			\centering
			\includegraphics[width=1.5in,height=1.3in]{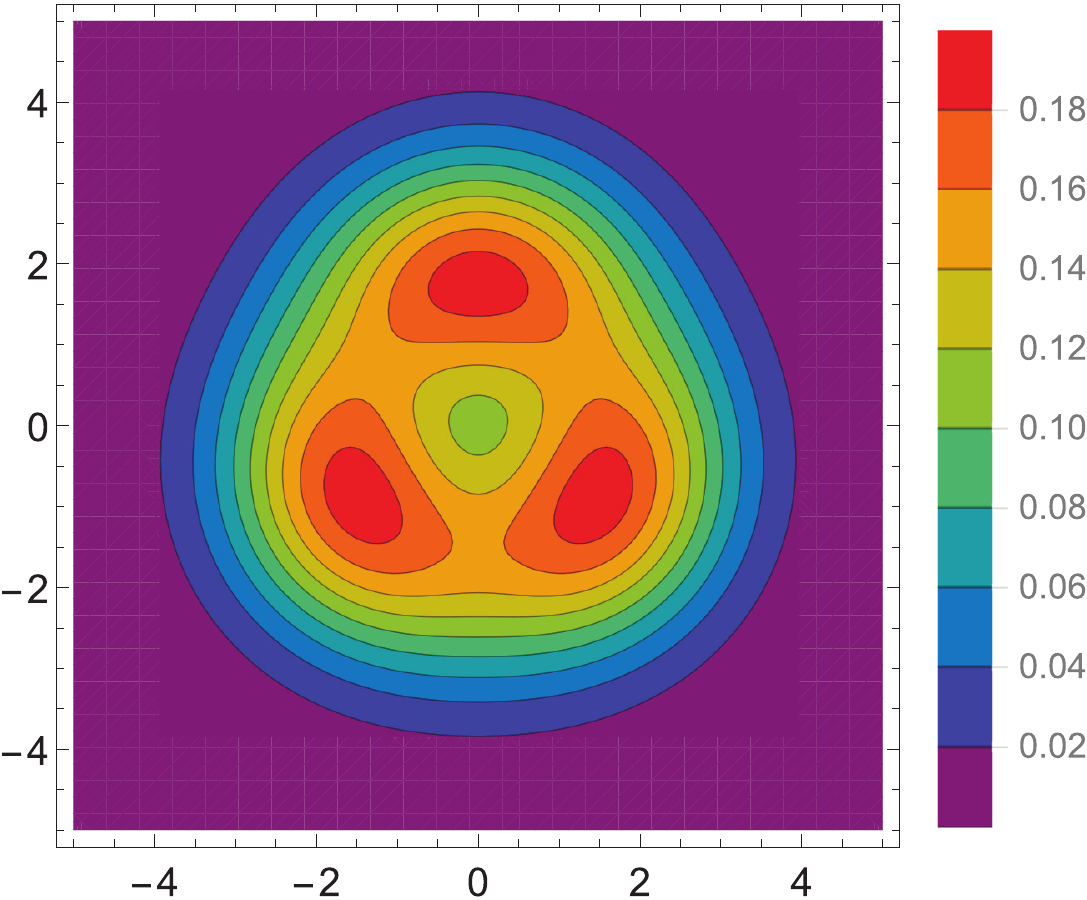}
		\end{minipage}
	}%
	\centering
	\caption{\label{fig:density} Various density profiles from the THSR wave functions with different shift parameters before angular momentum and parity projections in the two-dimensional container picture. 
		The size parameter is taken as $\beta_x=\beta_y=2.0$ fm, $\beta_z=0.5$ fm. The squared overlaps between the corresponding projected THSR wave functions $\Phi^{J^\pi}(S_{1x},S_{2x},S_{2y})$ and the Brink-GCM wave function are given. Figures share the same legend.}
\end{figure}

The obtained high accuracy of the THSR wave functions provides us with one clear nonlocalized cluster picture for understanding the intrinsic cluster structures of \ccg. According to the obtained optimal THSR wave functions in Fig.~\ref{fig:neg3k3sxy05pic} and Fig.~\ref{fig:neg4k3sxy05pic}, we take $\beta_x=\beta_y$=2.0 fm and $\beta_z$=0.5 fm as the size parameters for the common intrinsic wave functions of $3^-$ and $4^-$ states in Eq.~(\ref{hyb}). Thus, the THSR wave function $\Phi(\vect{\beta},\vect{S}_1,\vect{S}_2)$ in Eq.~(\ref{hyb}) can be rewritten as $\Phi(S_{1x},S_{2x},S_{2y})$ with the tacit understanding that the size parameters are those  from above. Various triangular shapes can then be constructed by taking the shift parameters $\vect{S}_1$ and $\vect{S}_2$ in the $xy-$plane. It can be seen that this adopted intrinsic THSR wave function has an oblate shape and it only slightly increases the widths of relative wave function in the $xy$ direction and almost makes no change in the $z$ direction compared with the Brink cluster model. Therefore, the  actual situation can be considered as the three $\alpha$'s moving in a two-dimensional container.
Figure~\ref{fig:density} shows the densities of various intrinsic wave functions with different values of shift parameters. It should be noted that Figure~\ref{fig:density}~(a) shows almost a positive-parity character of intrinsic density because of the "infinitesimal shifted triangle". To visualize the intrinsic structure of 3$\alpha$ clusters, we, therefore, take different lengths and orientations of the shift parameters as shown in Fig.~\ref{fig:density}~(b,c,d). It can be seen that, in spite of various configurations or geometrical shapes, their corresponding projected wave functions are quite similar and they are almost equivalent to the corresponding $3^-$ and $4^-$ GCM wave functions. It seems that the $\alpha$ clusters can almost move without restriction in our two-dimensional container picture what is quite different from the traditional localized, crystal-like, cluster picture. Moreover, it is worth realizing that, due to the Pauli principle, an effective localized clustering in the container model was found in the two-cluster \neneg\ system and 3$\alpha$ and 4$\alpha$ one-dimensional linear-chain system~\cite{suhara_onedimensional_2014a}. However, this is due to the constrained one-dimensional motion of the clusters. Here, in the present two-dimensional container, it seems that the 3$\alpha$ clusters feel little influence from the Pauli principle and show a real nonlocalized cluster motion. Because of the increased dimensionality, this is, of course, understandable, since in two dimensions it is much easier to get for the three $\alpha$'s out of one another's way. We think that this is the most important and rather spectacular finding of this work.

As we know, by using the Brink-GCM wave functions, the energy levels, electric transitions and some other observables of \ccg\ have been well reproduced~\cite{uegaki_structure_1979,kamimura_transition_1981}. Now, we found that these Brink-GCM wave functions are actually almost equivalent to our single THSR wave functions for the $3^-$ and $4^-$ states.
This successful description of the $3^-$ and $4^-$ states can be considered as our first attempt to study the negative-parity states of 3$\alpha$ clusters in the new container model. The introduced shift parameter can not only handle the negative-parity problem, but also provides us with a useful way to understand the nonlocalized cluster motion. On the other hand, there remain other negative-parity states whose structure stays so far unexplained.
In particular, as mentioned at the beginning, we found that the first $1^-$ and $5^-$ states cannot be well reproduced in our relatively simple picture of  single THSR wave functions which indicates that these $1^-$ and $5^-$ states have some special characters compared with the $3^-$ and $4^-$ states. Quite recently, the $1^-$ state at 10.8 MeV was reported~\cite{imai_realtime_2018} to have a strongly enhanced dipole transition strength from the ground state and could be regarded as an excitation mode of the Hoyle state. And the newly observed $5^-$ state with the high-excitation energy 22.4 MeV~\cite{marin-lambarri_evidence_2014} was actually difficult to be reproduced even with the GCM calculations. Therefore, in the next future step, it is expected that by including two $\alpha$ correlations or superposing enough configurations of the THSR-type wave function, these $1^-$ and $5^-$ states can be clarified in our model.

In summary, we successfully described the negative-parity $3^-$ and $4^-$ states in a simple container model. The obtained single optimal THSR wave functions for the $3^-$ and $4^-$ states have as high as 96\% and 92\% squared overlap with the corresponding GCM wave functions, respectively. From the densities of their intrinsic cluster wave functions, the nonlocalized cluster motion in the two-dimensional container was very clearly demonstrated.
\begin{acknowledgments}
This work was supported by JSPS KAKENHI Grant Numbers 17K14262 (Grant-in-Aid for Young Scientists (B)). Numerical computation in this work was carried out at the Yukawa Institute Computer Facility in Kyoto University.
\end{acknowledgments}
\newcommand{\noopsort}[1]{}

\end{document}